\documentclass[12pt,preprint]{aastex}
\usepackage{amsmath}
\newcommand{\ang}{\text{\normalfont\AA}}

\shorttitle{The Radio Quiet Quasar at $z=7.085$}
\shortauthors{Momjian et al.}

\begin{document}

\title{The Highest Redshift Quasar at $z=7.085$: A Radio Quiet Source}

\author{E. Momjian, C. L. Carilli}
\affil{National Radio Astronomy Observatory, P. O. Box O, Socorro, NM, 87801, USA}
\email{emomjian@nrao.edu,ccarilli@nrao.edu}

\author{F. Walter, B. Venemans}
\affil{Max-Planck-Institut fur Astronomie}
\email{walter@mpia.de,venemans@mpia.de}

\begin{abstract}
We present 1--2\,GHz Very Large Array A-configuration continuum observations on the highest redshift quasar known to date, the
$z=7.085$ quasar ULAS~J112001.48$+$064124.3. The results show no radio continuum emission at the optical position of the quasar or its
vicinity at a level of $\geq 3\sigma$ or $23.1~\mu$Jy~beam$^{-1}$. This $3\sigma$ limit
corresponds to a rest frame 1.4~GHz luminosity density limit of 
$L_{\rm \nu,1.4\,GHz} < 1.76 \times 10^{24}$~W~Hz$^{-1}$ 
for a spectral index of $\alpha=0$, and
$L_{\rm \nu,1.4\,GHz} < 1.42 \times 10^{25}$~W~Hz$^{-1}$ 
for a spectral index of $\alpha=-1$. The rest-frame
1.4~GHz luminosity limits are $L_{\rm rad} < 6.43 \times 10^6~L_{\odot}$ and $L_{\rm rad} < 5.20 \times 10^7~L_{\odot}$
for $\alpha=0$ and  $\alpha=-1$, respectively.
The derived limits for the ratio of the rest frame 1.4\,GHz luminosity density to the $B$-band optical
luminosity density are
$R\rlap{}_{1.4}^{*} < 0.53$ and $< 4.30$ for the above noted spectral indices, respectively.
Given our upper limits on the radio continuum emission and the radio-to-optical luminosity ratio, we conclude that this quasar
is radio-quiet and located at the low end of the radio quiet distribution of high redshift ($z \gtrsim 6$) quasars.

\end{abstract}


\section{INTRODUCTION}

The Sloan Digital Sky Survey (SDSS) revolutionized our understanding of the very high redshift Universe through the identification
of luminous quasars at $z \ge 6$ \citep{Fan00}. In the subsequent decade, some 50 quasars have been
discovered at $z > 5.7$. These high redshift quasars are proving to be critical probes of the tail-end of cosmic
reionization \citep{Fan06}, through studies of the Gunn-Peterson effect in their rest-frame UV spectra.
Observations in the cm and submm have also shown that a substantial fraction of these sources ($\sim 30\%$) are extreme
starburst galaxies, with  star formation rates $\sim 10^3$\,$M_\odot$ yr$^{-1}$, and molecular gas masses
$\sim 10^{10}$\,$M_\odot$ \citep{Wan10,CW13}. Hence, these systems are beacons to the coeval formation of supermassive
black holes and their host galaxies within 1\,Gyr of the Big Bang \citep{Wan10,Wan13}.

A curious characteristic of quasars is that their UV through X-ray spectra generally have properties indistinguishable
across cosmic time \citep{Fan04,She06,dr11}. This puzzling result suggests that the most distant quasars establish
their main emission characteristics quickly, even when the Universe was only 7\% of its current age. One recent discovery
that a change may occur at the highest redshifts is the detection of quasars with no emission from hot dust ($\sim 1000$\,K),
likely associated with the accretion disk and always seen in lower redshift quasar spectra \citep{Jia10}. A second
change in quasar characteristics is a possible decrease in the radio-loud fraction at the highest redshifts, which could
indicate a change in the relative accretion modes and spin for the first quasars \citep{Jia07,Wan07,Dot13}

In this paper, we extend our deep radio imaging to the highest redshift quasar known, ULAS~J112001.48$+$064124.3 (hereafter
J1120$+$0641) at $z = 7.085$.
This quasar was discovered in the
United Kingdom Infrared Telescope (UKIRT) Infrared Deep Sky
Survey (UKIDSS), and has an absolute magnitude of $M_{1450} = -26.6\pm0.1$ \citep{Mor11}.
The black hole powering the quasar has a mass of
$(2-3) \times 10^9$ $M_{\odot}$ (\citealt{Mor11}, De Rosa et al., in prep.), which is comparable to that of high-redshift
quasars found in the SDSS. Observations with the Plateau de Bure
Interferometer (PdBI) revealed a highly significant ($> 9\sigma$) detection of the [CII]
cooling line in the host galaxy of the quasar \citep{Ven12}. The underlying
far-infrared (FIR) continuum was also detected, and depending on the assumed properties of the dusty
spectral energy distribution (SED), it implies a FIR luminosity of
$L_\mathrm{FIR}=5.8\times10^{11}-1.8\times10^{12}$\,$L_{\odot}$ and a total dust mass in the
host galaxy of $6.7\times10^7-5.7\times10^8$\,$M_{\odot}$.

We compare our results on the $z=7.085$ quasar J1120$+$0641 to the well studied $z \sim 6$ quasar sample of \citet{Wan07}.
Throughout this paper, we assume a concordance cosmology with
$\Omega_{m}=0.27$, $\Omega_\Lambda=0.73$, and
${H_{0}=71}$~km~s$^{-1}$~Mpc$^{-1}$.

\section{OBSERVATIONS AND DATA REDUCTION}

The Observations of the $z=7.085$ quasar J1120$+$0641 were carried out with the Karl G. Jansky Very Large Array (VLA) of the NRAO\footnote{The National Radio Astronomy Observatory is a facility of the National Science Foundation operated under
cooperative agreement by Associated Universities, Inc.}
on 2011 August 9, 14, and 22, for a total of 4 hours. The array was in A-configuration (maximum baseline
= 36.4\,km).
The observations spanned the frequency range 1--2\,GHz. The WIDAR correlator was configured to
deliver 16 adjacent sub-bands in dual polarization, each 64~MHz wide and 128 spectral channels.

The calibrator 3C\,286 (J1331$+$3030) was used to set the absolute flux density scale and calibrate the
bandpass response, and the compact source J1120$+$1420 was observed as the complex gain calibrator.
Data editing, calibration, imaging, and analysis were performed using the Astronomical Image Processing System (AIPS;
\citealt{Gri03}) of the NRAO. 

Each observing session was edited and calibrated separately. The data sets were then concatenated,
imaged, and deconvolved. Imaging and deconvolution were performed through multi-faceting (200 fields)
to properly account for all the continuum sources in an area that covers $1.45$ square degrees centered at the location
of the $z=7.085$ quasar J1120$+$0641.

\section{RESULTS}

Fig.~1 shows the 1--2\,GHz continuum image centered at the optical location of the $z=7.085$ quasar J1120$+$0641.
The restoring beam size is  $1\rlap{.}{\arcsec}29 \times 1\rlap{.}{\arcsec}17$ (P.A.~$-57{\degr}$).
The contour levels are at $-$3, $-$2, 2, 3 times the rms noise level, which is 7.7\,$\mu$Jy~beam$^{-1}$. The image
was made with a grid weighting intermediate between pure natural and pure uniform (ROBUST = 0 in AIPS task IMAGR).

As seen in Fig.~1, no radio continuum emission is detected at the optical position of the quasar or in its vicinity.
This indicates the absence of any radio continuum emission with flux densities of $\geq 3\sigma$ or
$23.1~\mu$Jy~beam$^{-1}$.

Assuming a power-law ($S_{\nu} \propto \nu^{\alpha}$) radio spectral energy distribution, the 
$3\sigma$ limit on the radio continuum emission corresponds to a rest frame 1.4~GHz luminosity density limit of
$L_{\rm \nu,1.4\,GHz} < 1.76 \times 10^{24}$~W~Hz$^{-1}$ (or $< 1.76 \times 10^{31}$~ergs~s$^{-1}$~Hz$^{-1}$)
for a spectral index of $\alpha=0$,
and
$L_{\rm \nu,1.4\,GHz} < 1.42 \times 10^{25}$~W~Hz$^{-1}$ (or $< 1.42 \times 10^{32}$~ergs~s$^{-1}$~Hz$^{-1}$)
for a spectral index of $\alpha=-1$.
The resulting rest-frame
1.4~GHz luminosity limits are $L_{\rm rad} < 6.43 \times 10^6~L_{\odot}$ and $L_{\rm rad} < 5.20 \times 10^7~L_{\odot}$
for $\alpha=0$ and  $\alpha=-1$, respectively.

\section{DISCUSSION}

In the following we compare the measured parameters of the $z=7.085$ quasar J1120$+$0641
with the other high redshift ($z \sim 6 $) quasars discussed by \citet{Wan07}.

The absolute magnitude of J1120$+$0641 is $M_{1450} = -26.6 \pm0.1$ \citep{Mor11}. Following \citet{Wan07}, and assuming a
spectral index of $-0.5$, this results in a luminosity density at rest frame 4400\,\ang~of $L_{\nu,4400\,{\rm {\ang}}} = (3.3
\pm 0.3) \times 10^{24}$\,W\,Hz$^{-1}$, and a $B$-band luminosity of $L_{B}= \nu L_{\nu,4400\,{\rm {\ang}}}= (5.9 \pm 0.6) \times 10^{12}$\,$L_{\odot}$.

The radio loudness or quietness of a source can be investigated using the ratio of the observed radio to the
optical flux densities, $R$ \citep{sch70}. For instance, a value of $R \sim 10$ separates the
radio-loud and the radio-quiet sources for radio observations at 5\,GHz and optical observations at 4400\,{\rm {\ang}}
\citep{Kel89}. The K-corrected ratio of radio to optical power for given rest-frame radio frequency of $\nu$, $R\rlap{}_{\nu}^{*}$,
provides a more consistent
estimate among quasars of different redshifts \citep{SW80,sto92}. For the $z=7.085$ quasar, and using the
3$\sigma$=23.1\,$\mu$Jy\,beam$^{-1}$ noise level, the derived
limit for the ratio of the rest frame 1.4\,GHz luminosity density to the $B$-band optical
luminosity density, $R\rlap{}_{1.4}^{*}= L_{\rm \nu,1.4\,GHz} /L_{\nu,4400\,{\rm {\ang}}}$ is
$R\rlap{}_{1.4}^{*} < 0.53$ for $\alpha=0$ and $R\rlap{}_{1.4}^{*} < 4.30$ for $\alpha=-1$. 
The values of $R\rlap{}_{1.4}^{*}$ that separate radio-loud and the radio-quiet sources are
$\sim 10$ and $\sim 40$ for $\alpha=0$ and  $\alpha=-1$, respectively. Therefore, irrespective of the
assumed spectral index, the source J1120$+$0641 at $z=7.085$ is a radio quiet quasar.

Hereafter, and for proper comparison with the results obtained on other high redshift quasars ($z \sim 6$; \citealt{Wan07}),
we adopt their radio spectral index of $\alpha=-0.75$, and use the 2$\sigma$ noise level of our VLA
observations, which is 15.4\,$\mu$Jy\,beam$^{-1}$. This spectral index is appropriate because the majority ($\sim 85\%$) of
extragalactic radio sources
exhibit steep radio spectral slopes (e.g., \citealt{ki08} and references therein).
Furthermore, such a value is appropriate to characterize the synchrotron emission at low frequencies
from a normal galaxy, i.e., a galaxy with a radio energy source that is not a supermassive black hole \citep{condon92}. 
\citet{Wan07} have adopted this value to investigate the star formation properties of the host galaxies 
of $z \sim 6$ quasars.
For the above noted assumptions, the rest frame 1.4~GHz luminosity density limit of the $z=7.085$ quasar J1120$+$0641 is
$L_{\rm \nu,1.4\,GHz} < 5.62 \times 10^{24}$~W~Hz$^{-1}$, and the rest-frame
1.4~GHz luminosity limit is $L_{\rm rad} < 2.06 \times 10^7~L_{\odot}$.

To further investigate the radio--quietness of J1120$+$0641, we show the logarithm of its rest-frame 1.4\,GHz luminosity
($L_{\rm rad}$) vs.~the logarithm of its $B$-band luminosity ($L_{B}$) in Fig.~2. In this figure we also show the
$z \sim 6$ quasars studied by \citet{Wan07}.
The corresponding upper limit for $R\rlap{}_{1.4}^{*}= L_{\rm \nu,1.4\,GHz} /L_{\nu,4400\,{\rm {\ang}}}$ is
$R\rlap{}_{1.4}^{*} < 1.70$. For $\alpha=-0.75$, a value of $R\rlap{}_{1.4}^{*} \sim 30$ would separate the radio-loud and the
radio-quiet sources \citep{Wan07}. The $R\rlap{}_{1.4}^{*}$ of the quasar J1120$+$0641 is similar to the majority of the $z \sim 6$
quasars shown in Fig.~2. Overall, the sample of the high redshift quasars shown in Fig~2 is consistent with the
trend of deceasing radio-loud fraction with increasing redshift \citep{Jia07}.

The mass of the black hole powering this $z=7.085$ quasar is $(2-3) \times 10^9$ $M_{\odot}$ (\citealt{Mor11}, De Rosa et al.,
in prep). 
\citet{Dot13} studied the relationship between supermassive black holes in galactic nuclei and their spins.
These authors concluded that black holes with masses $M_{\rm BH} \sim 10^9\,M_{\odot}$ can have either low or high spins
depending on the fueling conditions. If the fueling of the black hole is not completely isotropic, then powerful radio jets
that are stable in their orientation would be produced. However, if the black hole has
been fed by random and isotropically distributed gas clouds, it will have lower spins and lower radio power.
Therefore, the radio-quietness of the  $z=7.085$ quasar is consistent with a very massive black hole with lower
spin due to isotropic accretion.

In the following we discuss the radio--FIR relation: Fig.~3 shows the logarithm of the rest-frame 1.4\,GHz
luminosity density ($L_{\rm \nu, 1.4\,GHz}$) vs.~the logarithm of the FIR luminosity ($L_{\rm FIR}$) range
of J1120$+$0641, which is $5.8\times10^{11}-1.8\times10^{12}$\,$L_{\odot}$ and 
indicated by the horizontal line, along with other quasars at $z \sim 6$ from \citet{Wan07}.
\citet{Yun01} quantified the ratio of the FIR and radio flux densities through the $q$ parameter,
\begin{equation}
q \equiv {\rm log}~\Big(\frac{\rm FIR}{3.75 \times 10^{12}~{\rm W\,m^{-2}}}\Big) -
{\rm log}~\Big(\frac{S_{\rm 1.4\,GHz}}{{\rm W\,m^{-2}\,Hz^{-1}}}\Big),
\end{equation}
and found $q=2.34$ for typical star-forming galaxies in the IRAS 2\,Jy sample. For reference, we also 
plot galaxies from the IRAS 2 Jy sample \citep{Yun01} with $L_{\rm FIR} \geq 10^{11.35} L_{\odot}$ in Fig.~3.
The lower limits for the $q$
parameter of the $z=7.085$ quasar using the 2$\sigma$ noise level of 15.4\,$\mu$Jy\,beam$^{-1}$ and the range of its FIR
luminosity are $q > 1.02 - 1.51$.

There are four quasars at $z \sim 6$ that have both radio and FIR detections (\citealt{Wan07}, and references therein, see
also Fig.~3). If the $q$ value of the $z=7.085$ quasar J1120$+$0641 were to be 
comparable to the average $q$ value of
these four quasars ($q = 1.72$), then the expected radio flux density at the frequency of our observations
would have been $\sim~3.1-9.5$\,$\mu$Jy\,beam$^{-1}$ ($0.4\sigma -1.2\sigma$ significance) for the range of its $L_{\rm FIR}$.
If, however, the  $z=7.085$ quasar were to follow the radio-FIR correlation ($q = 2.34$; \citealt{Yun01}), then the expected 
radio flux density would have been
 $\sim 0.7-2.3$\,$\mu$Jy\,beam$^{-1}$ ($0.1\sigma -0.3\sigma$ significance).
The upper limit for the radio flux density of the $z=7.085$ quasar J1120$+$0641 from our observations is not adequate to probe the
radio-FIR relationship in the context of star formation. Deeper radio continuum observations are needed to carry out this
type of studies, but would require about 100\,hours of integration time to go significantly (factor $\sim 5$) deeper than the
current limit.

\section{Conclusions}
We have presented 1--2\,GHz VLA A-configuration continuum observations on the highest redshift quasar known to date, the
$z=7.085$ quasar J1120+0641. The results show no radio continuum emission at the optical position of the quasar or its
vicinity at a level of $\geq 3\sigma$ or $23.1~\mu$Jy~beam$^{-1}$. For a range of assumed spectral indices of 0 to $-1$, we derive 
rest frame 1.4~GHz luminosity density limits of $L_{\rm \nu,1.4\,GHz} < 1.76 \times 10^{24} - 1.42 \times 10^{25}$~W~Hz$^{-1}$,
and rest frame 1.4~GHz luminosity limits of $L_{\rm rad} < 6.43 \times 10^6 - 5.20 \times 10^7~L_{\odot}$.
For the above noted range of spectral indices, the limits for the ratio of the rest frame 1.4\,GHz
luminosity density to the $B$-band optical luminosity density, $R\rlap{}_{1.4}^{*}$, are $<0.53 - 4.30$.
The derived values clearly show that this source is a radio-quiet quasar, and the comparison with other high redshift ($z
\sim 6$) quasars places it at the low end of the radio quiet distribution of such quasars.
We also derive lower limits for the $q$ parameter of the $z=7.085$ quasar J1120+0641. However, 
the sensitivity limits of our observations are not adequate to probe the star formation
in the host galaxy of this quasar and study the radio-FIR relationship.


\begin{figure}
\epsscale{1}
\plotone{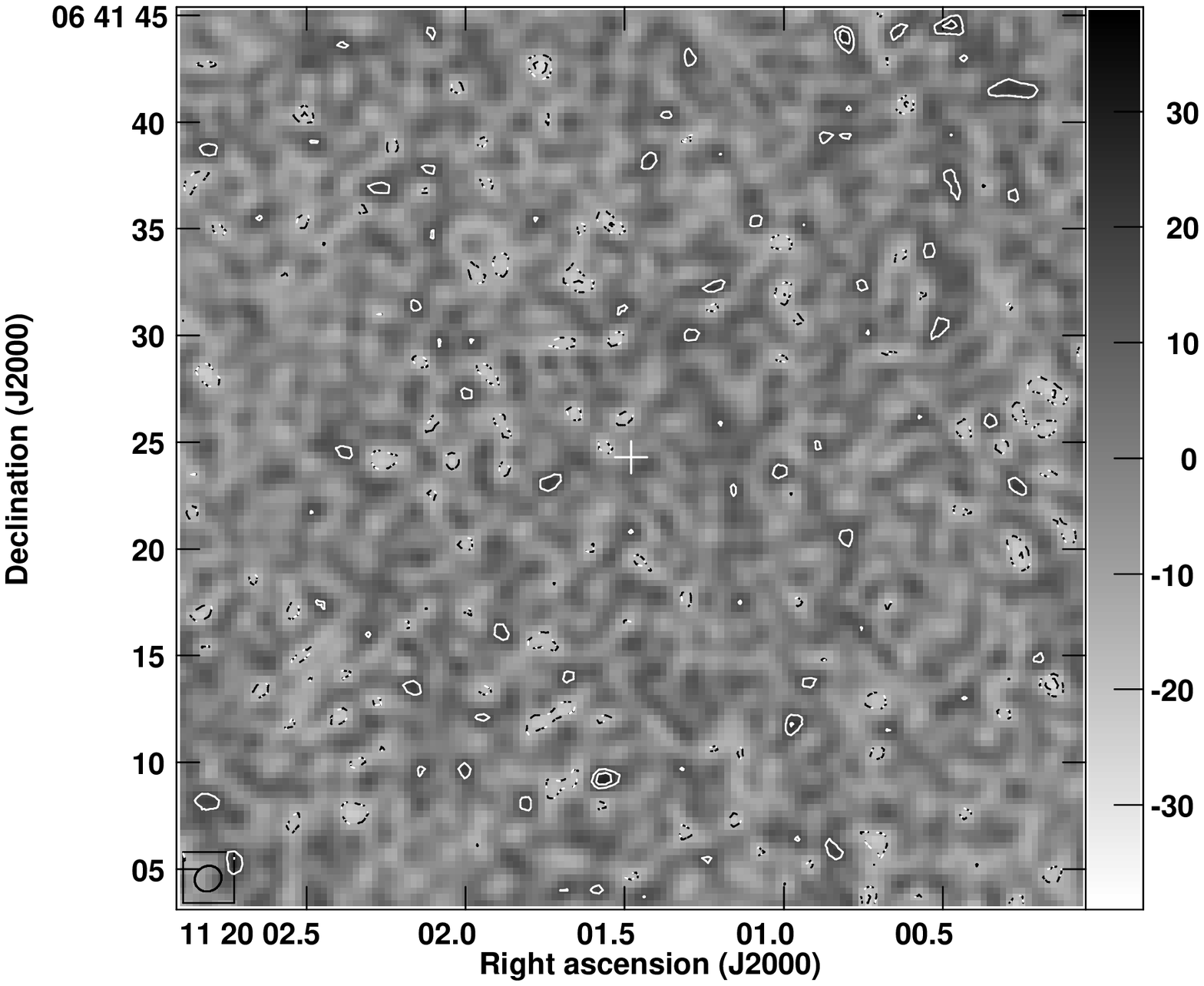}
\figcaption{
Continuum image (1--2\,GHz) centered at the optical location of the $z=7.085$ quasar J1120+0641.
The restoring beam size is $1\rlap{.}{\arcsec}29 \times 1\rlap{.}{\arcsec}17$ in position angle $-57{\degr}$, 
as shown in the lower left side of the figure.
The contour levels are at $-$3, $-$2, 2, 3 times the rms noise level,
which is 7.7\,$\mu$Jy~beam$^{-1}$. The gray-scale range is indicated
at the right side of the image in units of $\mu$Jy~beam$^{-1}$. 
The plus sign indicates the optical position of the quasar, which is 
$\alpha(\rm{J2000.0})= 11^{\rm h}20^{\rm m}01\rlap{.}^{\rm s}48$,
$\delta(\rm{J2000.0})=+06^{\circ}41{\arcmin}24\rlap{.}{\arcsec}3$ \citep{Mor11}.
\label{f1}}
\end{figure}

\clearpage
\begin{figure}
\includegraphics[scale=0.8]{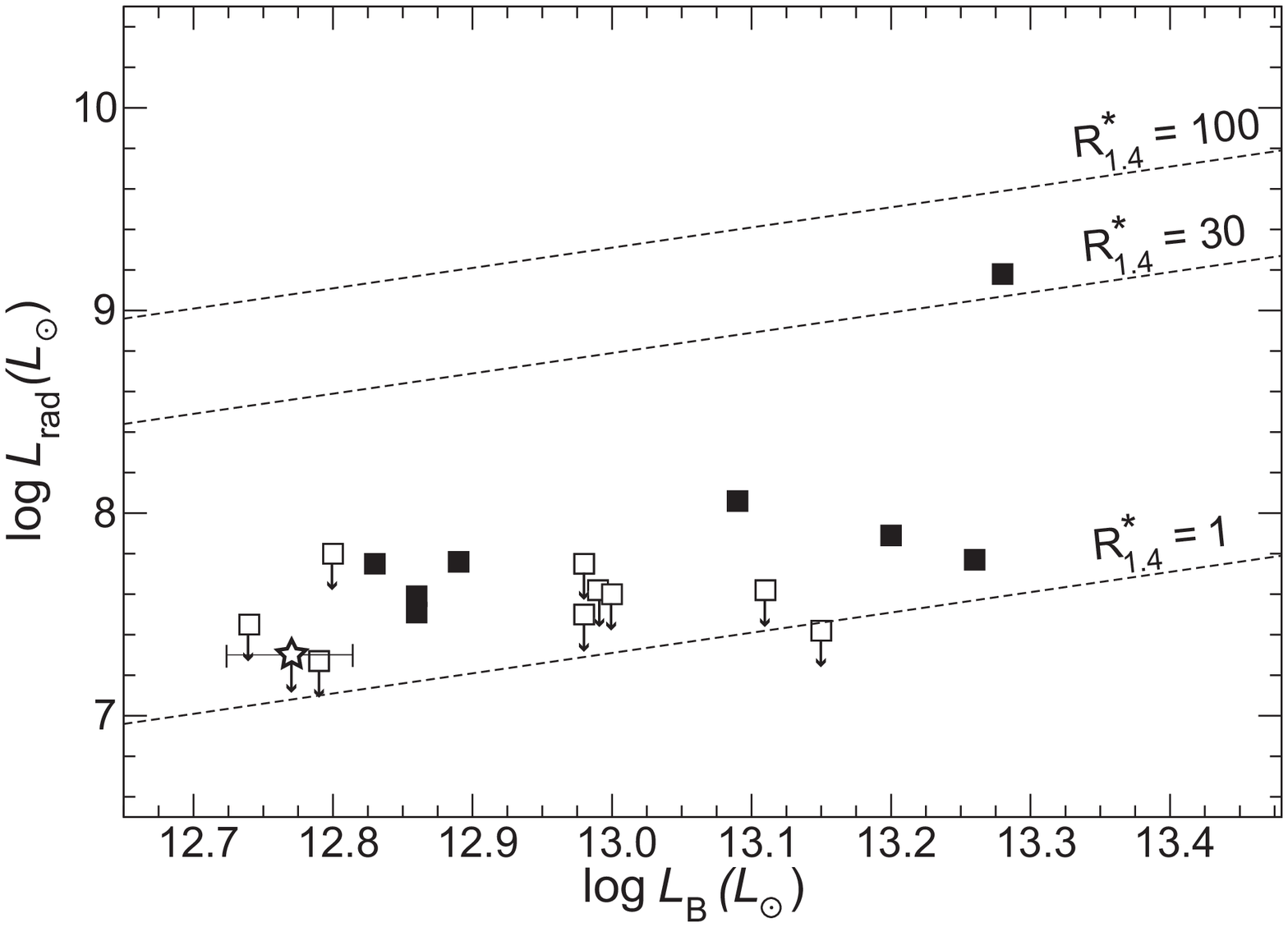}
\figcaption{
Logarithm of the rest-frame 1.4\,GHz luminosity vs. the logarithm of the $B$-band luminosity
of $z \gtrsim 6$ quasars.
The filled and open squares are detections and upper limits, respectively, from \citet{Wan07}. The star sign marks
the $z=7.085$ quasar J1120$+$0641 using the 2$\sigma$ rms noise limit of our VLA observations and assuming  a spectral index of $\alpha=-0.75$
for consistency with
the non-detection limits reported by \citet{Wan07}, and its $L_{B}$ value of $5.9 \times 10^{12}\,L_{\odot}$. The horizontal bar
reflects the error in $L_{B}$ (see \S 4).
The dashed lines represent rest-frame radio-to-optical ratios $R\rlap{}_{1.4}^{*}$ of 1, 30, and 100.
The $R\rlap{}_{1.4}^{*} \sim 30$ denotes the separation of radio-loud and radio quiet quasars for the assumed spectral index \citep{Wan07}.
\label{f2}}
\end{figure}

\clearpage
\begin{figure}
\includegraphics[scale=0.8]{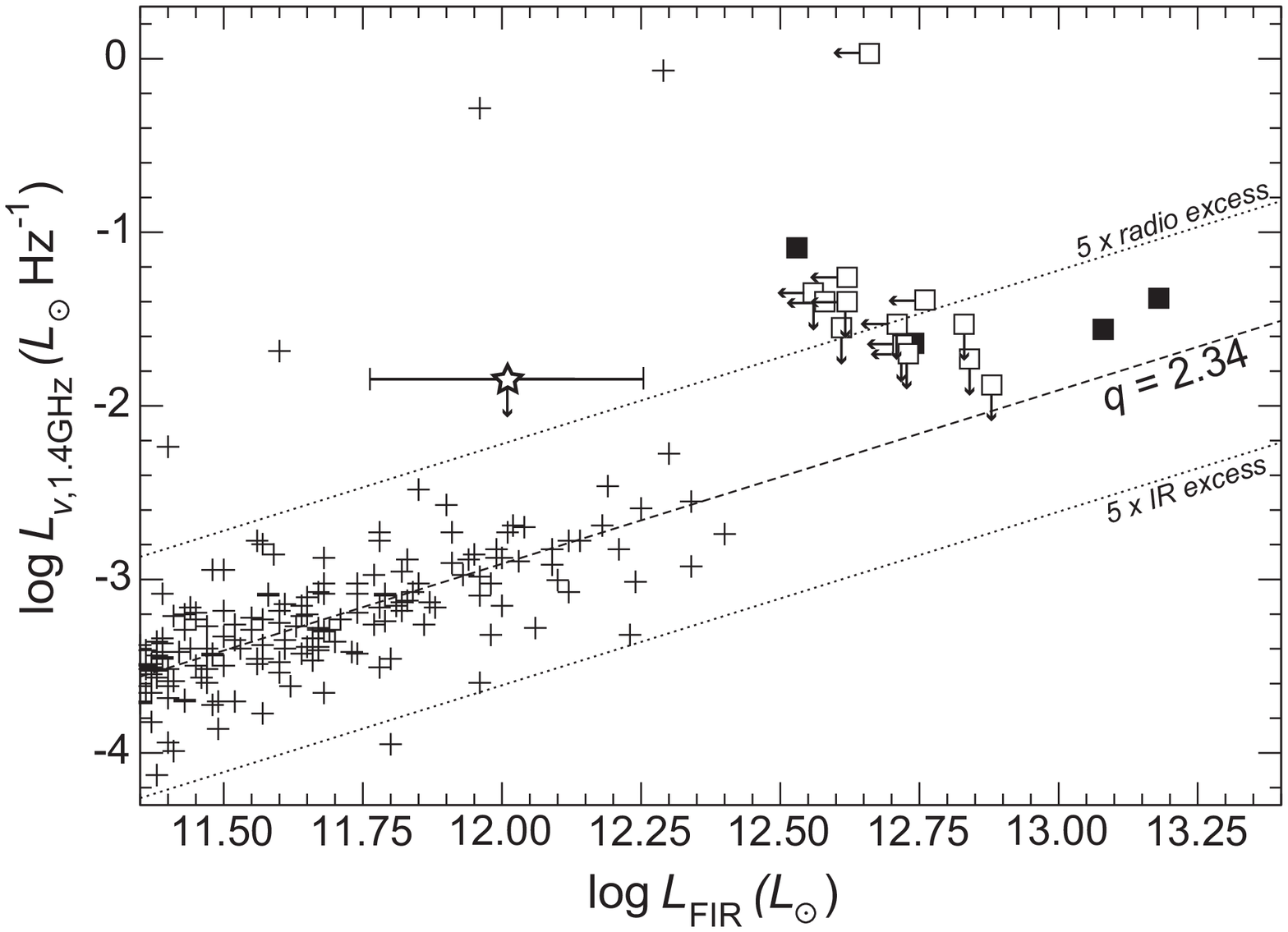}
\figcaption{
Logarithm of the rest-frame 1.4 GHz luminosity density vs. the logarithm of the FIR luminosity
of $z \gtrsim 6$ quasars.
The filled and open squares are detections and upper limits, respectively, from \citet{Wan07}.
The star sign marks
the $z=7.085$ quasar J1120$+$0641 using the 2$\sigma$ rms noise limit of our VLA observations and assuming  a spectral index of $\alpha=-0.75$ for consistency with
the non-detection limits reported by \citet{Wan07}, and its mean $L_{\rm FIR}$ value of $1.19 \times 10^{12}\,L_{\odot}$.
The horizontal bar reflects the range of the $L_{\rm FIR}$ values for this quasar (see \S 1).
The plus signs represent the IRAS 2 Jy sample of galaxies in \citet{Yun01}.
The dashed line indicates the typical radio-to-FIR correlation in star-forming galaxies with a correlation parameter
$q = 2.34$ \citep{Yun01}. The dotted lines represent the radio-excess (above) and IR-excess (below)
objects that have 5 times larger radio and FIR flux density than the expected value from the linear radio-FIR correlation, respectively.
\label{f3}}
\end{figure}

\end{document}